\documentclass[aps,prb,twocolumn,
			   groupedaddress,superscriptaddress,
			   amsfonts,amssymb,amsmath, 
			   citeautoscript,
			   a4paper
			   ]{revtex4-1}

\usepackage{txfonts}  %Times Roman fonts
\usepackage{txfontsb} %Addition for txfonts, including old style numerals and greek
\usepackage{bm} %Bold math symbols with \bm{} (Greek and other symbols)
\usepackage{microtype} %For better kerning and symbol-stretching

\usepackage{xcolor}
\usepackage{graphicx}
\usepackage{tikz}

\usepackage{hyperref}
\hypersetup{
    colorlinks,
    linkcolor={blue!75!black!80!yellow},
    citecolor={blue!75!black!80!yellow},
    urlcolor={blue!75!black!80!yellow}
}

\usepackage[centering,hmargin=16mm,tmargin=30mm,bmargin=26mm]{geometry}

\usepackage{enumitem}

\newcommand{\rv}{\mathbf{r}}
\newcommand{\Ev}{\mathbf{E}}
\newcommand{\ef}{\epsilon_{\text{\textsc{f}}}}
\newcommand{\vf}{v_{\text{\textsc{f}}}}

\newcommand{\e}{\mathrm{e}}
\newcommand{\dd}[1]{\mathrm{d}#1\,}
\newcommand{\appropto}{\mathrel{\vcenter{
  \offinterlineskip\halign{\hfil$##$\cr
    \propto\cr\noalign{\kern.2pt}\sim\cr\noalign{\kern-2.5pt}}}}}
\newcommand{\kpar}{k_{\scriptscriptstyle\parallel}}

\makeatletter
\renewcommand{\fnum@figure}{\figurename~\thefigure\ (color online)}
\makeatother

\usepackage{titlesec}

\begin{document}
%-----------------
%----- TITLE -----
%-----------------
\title{Kerr nonlinearity and plasmonic bistability in graphene nanoribbons}

%------------------------------------
%----- AUTHORS AND AFFILIATIONS -----
%------------------------------------

\author{Thomas~Christensen}
\affiliation{Department of Photonics Engineering, Technical University of Denmark, DK-2800 Kgs. Lyngby, Denmark}
\affiliation{Center for Nanostructured Graphene, Technical University of Denmark, DK-2800 Kgs. Lyngby, Denmark}
\author{Wei~Yan}
\affiliation{Department of Photonics Engineering, Technical University of Denmark, DK-2800 Kgs. Lyngby, Denmark}
\affiliation{Center for Nanostructured Graphene, Technical University of Denmark, DK-2800 Kgs. Lyngby, Denmark}
\author{Antti-Pekka~Jauho}
\affiliation{Center for Nanostructured Graphene, Technical University of Denmark, DK-2800 Kgs. Lyngby, Denmark}
\affiliation{Department of Micro- and Nanotechnology, Technical University of Denmark, DK-2800 Kgs. Lyngby, Denmark}
\author{Martijn~Wubs}
\affiliation{Department of Photonics Engineering, Technical University of Denmark, DK-2800 Kgs. Lyngby, Denmark}
\affiliation{Center for Nanostructured Graphene, Technical University of Denmark, DK-2800 Kgs. Lyngby, Denmark}
\author{N.~Asger~Mortensen}
\email{asger@mailaps.org}
\affiliation{Department of Photonics Engineering, Technical University of Denmark, DK-2800 Kgs. Lyngby, Denmark}
\affiliation{Center for Nanostructured Graphene, Technical University of Denmark, DK-2800 Kgs. Lyngby, Denmark}

%---------------------------
%----- KEYWORDS & PACS -----
%---------------------------

%--------------------
%----- ABSTRACT -----
%--------------------
\begin{abstract}
We theoretically examine the role of Kerr nonlinearities for graphene plasmonics in nanostructures, specifically in nanoribbons. The nonlinear Kerr interaction is included semiclassically in the intraband approximation. The resulting electromagnetic problem is solved numerically by self-consistent iteration with linear steps using a real-space discretization. We derive a simple approximation for the resonance shifts in general graphene nanostructures, and obtain excellent agreement with numerics for moderately high field strengths. Near plasmonic resonances the nonlinearities are strongly enhanced due to field enhancement, and the total nonlinearity is significantly affected by the field inhomogeneity of the plasmonic excitation. Finally, we discuss the emergence of a plasmonic bistability which exists for frequencies redshifted relative to the linear resonance. 
Our results offer new insights into the role of nonlinear interaction in nanostructured graphene and paves the way for experimental investigation.
\end{abstract}

%---------------------
%----- MAKETITLE -----
%---------------------
\maketitle

%------------------------
%----- INTRODUCTION -----
%------------------------
Nonlinear optical effects~\cite{Boyd:2008,Gibbs:1985}, facilitated by strong light-matter interaction, are indispensable in modern photonics. Indeed, a host of phenomena and applications arise at sufficiently high field-strengths, owing to superlinear photon-photon response mediated by strong light-matter interaction, ranging from frequency conversion, through all-optical phase-modulation, to ultra-fast switching, and is pursued in a broad range of platforms~\cite{Soljacic:2004,Dudley:2009,Leuthold:2010}. 

A perennial challenge in the discipline is to achieve significant nonlinear interaction at ever smaller excitation powers and interaction volumes, whilst maintaining in-situ tunability and control. In achieving this goal, the field of plasmonics, describing the strong hybridization of the free electromagnetic field with collective oscillations of conduction electrons, suggests several promising avenues~\cite{Kauranen:2012}. 
In particular, the extreme local field enhancements inherent to plasmonic excitations amplify intrinsic nonlinearities considerably, allowing large effective nonlinearities.

Nevertheless, plasmonic field-enhancement is fundamentally limited by intrinsic Ohmic losses even in noble metals. The advent of the two-dimensional material graphene has garnered significant interest in the plasmonic community~\cite{Bludov:2013,Abajo:2014}, in part due to extremely large electron mobilities~\cite{Novoselov:2004,Bolotin:2008,Tassin:2013} and concomitant extraordinary plasmonic field-enhancements~\cite{Thongrattanasiri:2013a}, exceeding even the very large enhancements known from metal-plasmonics. Furthermore, graphene has attracted much interest also for its exceptional intrinsic nonlinear properties both theoretically~\cite{Mikhailov:2008, Mikhailov:2012,Yao:2012, Cheng:2014} and experimentally~\cite{Hendry:2010,Zhang:2012,Gu:2012}. Building on this compound-fortuity, a body of research is rapidly emerging at the crossroad of nonlinear plasmonics and graphene~\cite{Gullans:2013,Yao:2014,Smirnova:2014,Gorbach:2013,Peres:2014,Cox:2014a,Cox:2015a}.

Very recently, the role of Kerr nonlinearities in infinitely extended graphene has been studied~\cite{Peres:2014}, notably establishing the existence of bistable solutions.
In this paper we study theoretically an analogous Kerr nonlinearity but in nanostructured graphene, specifically in nanoribbons -- wherein plasmons, unlike in the extended counterpart, are readily excited without momentum-matching concerns, e.g.\ by normally incident plane waves. 
We report an induced nonlinearity which is strongly affected by the degree of inhomogeneity of the electric fields of the plasmon -- a feature which is absent in the corresponding extended system. Furthermore, we derive a simple perturbative expression for the nonlinear resonance shifts in general graphene nanostructures, and show that it is in excellent agreement with full self-consistent calculations for moderately high field strengths. Finally, we discuss the emergence of plasmonic bistability in nanoribbons under normally incident plane-wave excitation. 
First, however, we introduce the two basic components needed for a nonlinear treatment of graphene nanostructures, namely a material response model and an exposition of the resulting electromagnetic problem.

%------------------
%----- THEORY -----
%------------------

\vskip .35em
\emph{Material response.}\ \ %
For photon energies $\hbar\omega$ low comparable with the Fermi energy $\ef$, the response of graphene is reasonably approximated by neglecting interband transitions. In this case, the intraband response can be derived from the Boltzmann equation. To third order in the perturbing field the Kerr-corrected conductivity, i.e.\ the response oscillating at the perturbing frequency, is~\cite{Peres:2014}
\begin{equation}\label{eq:Kerr}
\sigma(\rv) = \sigma_{\scriptscriptstyle(1)}\Big[1 - |\Ev(\rv)|^2/E_{\scriptscriptstyle(3)}^2\Big],
\end{equation}
expressed in terms of the linear intraband conductivity $\sigma_{\scriptscriptstyle(1)} = \mathrm{i}e^2\ef/\pi\hbar^2(\omega+\mathrm{i}\gamma)$ with loss-rate $\gamma$, and a third-order characteristic field $E_{\scriptscriptstyle(3)}^2 \equiv (8\varpi_{\scriptscriptstyle(3)}^2)/(9\omega^2)E_{\text{sat}}^2$ linearly related to the saturation field $E_{\text{sat}}\equiv \ef\omega/e\vf$ through a loss-modified frequency $\varpi_{\scriptscriptstyle(3)}^2 \equiv (\omega+\tfrac{1}{2}\mathrm{i}\gamma)(\omega-\mathrm{i}\gamma)$. Since the Kerr correction is of the self-focusing type~\cite{Gorbach:2013}, its usage in finite structures with inhomogeneous fields must be augmented to include a saturating mechanism, or else suffer nonphysical run-away self-focusing~\cite{Wang:1995}. Here we adopt the well-known two-level saturation model, or, in other words, the $[0/2]$ Pad\'{e} approximant of $\sigma(\rv)$ consistent with Eq.~\eqref{eq:Kerr}
\begin{equation}\label{eq:saturation}
\sigma(\rv) \simeq \frac{ \sigma_{\scriptscriptstyle(1)}(\rv)}{1+|\Ev(\rv)|^2/E_{\scriptscriptstyle(3)}^2}
+\sigma_{\scriptscriptstyle(3)2\gamma}(\rv).
\end{equation}
This model reproduces the third-order result of Eq.~\eqref{eq:Kerr} in the $|\Ev(\rv)|/E_{\text{sat}}\ll 1$ limit, while crucially exhibiting a sensible behavior beyond this limit as well~\cite{Note1}.
Lastly, we include in Eq.~\eqref{eq:saturation} a term $\sigma_{\scriptscriptstyle(3)2\gamma}(\rv)$ to account for a high-field loss mechanism through two-photon absorption via the phenomenological prescription suggested by Gorbach~\cite{Gorbach:2013}, via the dissipative correction $\sigma_{\scriptscriptstyle(3)2\gamma}(\rv) = -\mathrm{i}\alpha_{\scriptscriptstyle 2\gamma}\sigma_{\scriptscriptstyle (1)}|\Ev(\rv)|^2/E_{\text{sat}}^2$ with $\alpha_{\scriptscriptstyle 2\gamma} \approx 0.1$ estimated from measurements~\cite{Gu:2012}.

Before proceeding we briefly discuss the limitations of the material response assumed in Eq.~\eqref{eq:saturation}. Firstly, the disregard of interband effects limits our consideration to energies sufficiently below $\sim\!\!2\ef$. Secondly, nonlocality~\cite{Wang:2013}, edge-states~\cite{Christensen:2014prb}, and more generally atomistic features~\cite{Cox:2014a, Cox:2015a,Wang:2015, Silveiro:2015} are excluded, though they are important at small feature sizes. Consequently, we restrict our considerations to nanostructures of characteristic dimensions $\gtrsim 25\ \text{nm}$ where these effects only weakly perturb the intraband approximation.

%-------------------------------------------------------------
\vskip .35em
\emph{Interacting response.}\ \ %
In the quasistatic limit, the interacting response of graphene can be deduced from three elements; the Coulomb law, the continuity equation, and the current-field relationship as specified by a conductivity-model. For a nanostructure defined by a two-dimensional domain $\Omega$ (e.g.\ at $z=0$), these elements combine to form an integro-differential equation for either the induced density or the total potential $\phi(\rv)$. Here we choose the latter~\cite{Abajo:2014}
\begin{equation}\label{eq:phi_integrodiff}
\phi(\rv) = \frac{\mathrm{i}}{4\pi\varepsilon_0\omega{}W} \int_{\Omega} \!\mathrm{d}^2{\rv'}\, V(\rv,\rv') \nabla'\!\cdot\big[\sigma(\rv')\nabla'\phi(\rv')\big],
\end{equation}
expressed in dimensionless coordinates $\rv^{(\prime)}=[x^{(\prime)},y^{(\prime)},z]^{\text{\textsc{t}}}$ normalized by a characteristic length $W$, with the Coulomb interaction $V(\rv,\rv') = |\rv-\rv'|^{-1}$, and with differential operators $\nabla' = [\partial_{x'},\partial_{y'}]^{\text{\textsc{t}}}$. The conductivity $\sigma(\rv)$ implicitly depends on frequency -- and in a nonlinear treatment also on the total field $\Ev(\rv)$. The spatial dependence of the conductivity can be conveniently expressed via a dimensionless occupation function $f(\rv)\equiv \sigma(\rv)/\langle\sigma_{\scriptscriptstyle (1)}\rangle$ with $\langle\sigma_{\scriptscriptstyle (1)}\rangle$ denoting the average linear conductivity across $\Omega$. Introducing operators $\textsf{V}g(\rv) \equiv \int \dd{\rv'} V(\rv,\rv') g(\rv')$ and $\textsf{D}g(\rv') \equiv \nabla'\!\cdot[f(\rv')\nabla'g(\rv')]$ casts Eq.~\eqref{eq:phi_integrodiff} as an eigenvalue problem for the composite operator $\textsf{V}\textsf{D}$
\begin{equation}\label{eq:eigsys}
 \lambda\phi(\rv) = \textsf{V}\textsf{D}\phi(\rv),
\end{equation}
with eigenvalues $\lambda\equiv 4\pi\varepsilon_0\omega W/\mathrm{i}\langle\sigma_{\scriptscriptstyle (1)}\rangle$, dictating the permitted eigenfrequencies $\omega$. Operators $\textsf{V}$ and $\textsf{D}$ find simple matrix-forms in a discretized real-space basis in both the general 2D case as well as in the 1D ribbon case, see Supplemental Material (SM). The operator $\textsf{D}$ is constructed so as to account explicitly for a boundary condition of vanishing normal current (or, equivalently, for the conductivity-discontinuity) at the boundary $\partial\Omega$. In the presence of an external potential $\phi_{\text{ext}}$, the eigenvalue problem in Eq.~\eqref{eq:eigsys} becomes an inhomogeneous equation through the addition to the right-hand-side of a source-term $\lambda\phi_{\text{ext}}(\rv)$.
To solve the nonlinear problem, with $\sigma(\rv)$, and hence $f(\rv)$ and $\textsf{D}$, depending on the total electric field locally, we solve the nonlinear system iteratively until self-consistency is reached, exploiting at each iteration-step the computational efficiency associated with linear systems~\cite{Wang:1995}, see SM.

%-----------------------
\begin{figure}[!htb]\center
\includegraphics[height=7.85cm]{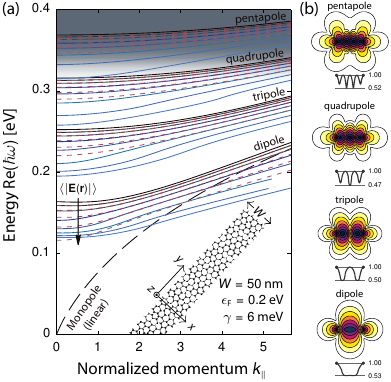}
\caption{\textsf{(a)} Dispersion relation of a single nanoribbon. Ribbon-averaged field strength $\langle|\Ev(\rv)|\rangle$ ranges from negligible (black), i.e. linear, through $1\times 10^5\ \text{V/cm}$ to $4\times 10^5\ \text{V/cm}$ (lightest blue) in steps of $0.5\times 10^5\ \text{V/cm}$ (increasing along arrow). For the first five $\langle|\Ev(\rv)|\rangle$, we indicate in dashed red the corresponding analytical estimate, see Eq.~\eqref{eq:perturb}. For the monopole, only the linear calculation is shown. The region of significant interband-modification is illustrated in shaded gray. Inset schematically depicts a single graphene nanoribbon. \textsf{(b)} Field intensity, $|\Ev(\rv)|$, contour maps for the case $\langle|\Ev(\rv)|\rangle=4\times 10^5\ \text{V/cm}$ and $\kpar=0$. Colormap ranges from maximal (dark) to minimal (light) logarithmically, with contours separated by factors of 1.5, 1.75, 2, and 2.25 for dipole, tripole, quadrupole, and pentapole cases, respectively. Sparklines below maps depict the variation of $|f(\rv)|$ along the ribbon, with maximal and minimal values indicated.}
\label{fig:disprel}
\end{figure}
%-----------------------

%--------------------------------
%----- RESULTS & DISCUSSION -----
%--------------------------------
With the formal premise established, we next specialize to the case of nanoribbons, translationally invariant along $y$ and of finite extent $W$ along $x$; a system which has already attracted much attention in the linear case~\cite{ChristensenJohan:2012,Wang:2013,Velizhanin:2015,Silveiro:2015}. As a consequence of translational symmetry, eigensolutions can be expanded in a momentum basis according to $\phi(\rv)= \phi(x,z)\exp(\mathrm{i} \kpar y)$. Of key interest is the evolution of the eigenenergies with momentum $\kpar$ (here dimensionless; conventional units via $\kpar/W$), i.e.\ the dispersion relation $\hbar\omega_n(\kpar)$ -- and subsequently the response of the system to external fields.

%----------------------------------------------------------------------
\vskip .35em
\emph{Eigenmodes and nonlinear dispersion.}\ \ %
For low field strengths, i.e.\ in the linear regime with $f(\rv)$ independent of $\Ev(\rv)$, the eigenmodes $\lambda_n(\kpar)$ of Eq.~\eqref{eq:eigsys} are solely geometry dependent -- but scale invariant -- with associated eigenfrequencies $\omega_n(\kpar)$ dictated by $\lambda_n(\kpar) = 4\pi\varepsilon_0\omega_n(\kpar) W/\mathrm{i}\langle\sigma_{\scriptscriptstyle (1)}\rangle$, allowing in the linear intraband approximation the simple scaling relation $\omega_n(\kpar)\simeq(2\pi)^{-1}\sqrt{-\lambda_n(\kpar)e^2\ef/ \varepsilon_0 W}$~\cite{ChristensenJohan:2012,Wang:2013}. Under significant nonlinear interaction, however, the eigenvalues $\lambda_n(\kpar)$ are field-dependent, and, by extension, scale-dependent due to the self-consistent nature of the problem. 
In Fig.~\ref{fig:disprel}(a) we investigate the dispersion relation of the first few eigenmodes of a single $W=50\ \text{nm}$ nanoribbon for different ribbon-averaged field strengths $\langle|\Ev(\rv)|\rangle \equiv W^{-1}\!\int_\Omega\dd{x}|\Ev(x)|$.
The most apparent impact of nonlinearity is a redshift of all resonances. This is readily appreciated from the negativity of the Kerr correction. Indeed, the shift can be well-approximated by perturbation theory for any general structure: denoting by $\hbar\omega_n^{\scriptscriptstyle(0)}$ and $\Ev_n^{\scriptscriptstyle(0)}$ the \emph{linear} response eigenenergies and eigenfields [with $\langle|\Ev_n^{\scriptscriptstyle(0)}(\rv)|\rangle = \langle| \Ev(\rv)|\rangle$] the nonlinear eigenenergies are, to lowest order, approximately (see SM)
\begin{equation}\label{eq:perturb}
\omega_n \simeq \omega_n^{\scriptscriptstyle (0)} \sqrt{
1 - \frac{9}{8}
\frac{\langle |\mathbf{E}^{\scriptscriptstyle (0)}(\rv)|^4\rangle}
{\langle |\mathbf{E}^{\scriptscriptstyle (0)}(\rv)|^2\rangle    E_{\text{sat}}^2}},
\end{equation}
with the averages taken over $\rv\in\Omega$. 
The approximation is excellent for moderately high fields, see dashed red lines of Fig.~\ref{fig:disprel}(a), though, naturally, inaccurate for the largest considered fields due to the disregard of the self-consistent aspects of the nonlinear perturbation. The approximation also reveals the important role played by the inhomogeneity of $\Ev(\rv)$, or equivalently $f(\rv)$, for the nonlinear strength since $\langle |\mathbf{E}^{\scriptscriptstyle (0)}(\rv)|^4\rangle \neq \langle |\mathbf{E}^{\scriptscriptstyle (0)}(\rv)|^2\rangle^2$ for inhomogeneous fields.

In Fig.~\ref{fig:disprel}(b) we explore this point further, by depicting the modal character and inhomogeneous nature of the plasmonic modes. The modal labels are chosen from the perspective of the induced charge density, $\rho(x)$, of the $n$th mode, with the monopole, dipole, tripole, quadrupole, and pentapole ($n=0,1,2,3,\text{ and }4$, respectively) exhibiting $n$ nodes of $\rho(x)$. Modes of even $n$ are optically dark, owing to a vanishing dipole moment, and remain optically dark also under nonlinear perturbations (which preserves the system symmetry). The monopole violates charge conservation along $x$ [but not along $(x,y)$ for $\kpar\neq0$], is optically dark, and consistently does not converge at higher fields; as a consequence, we depict only its linear dispersion. 
The variation of the occupation function $f(\rv)$ under large fields is highlighted in the insets of Fig.~\ref{fig:disprel}(b). The strong spatial variation of the conductive profile, up to 50\% for the considered $\langle|\Ev(\rv)|\rangle$, is a direct consequence of the strongly inhomogeneous nature of plasmons. 
Despite these significant spatial variations of $f(\rv)$, the corresponding modal profiles away from the ribbon are highly similar under linear and nonlinear circumstances, since their character is dictated chiefly by the nodal character of $\rho(x)$.

%-----------------
\begin{figure}[!htb]\center
\includegraphics[scale=.925]{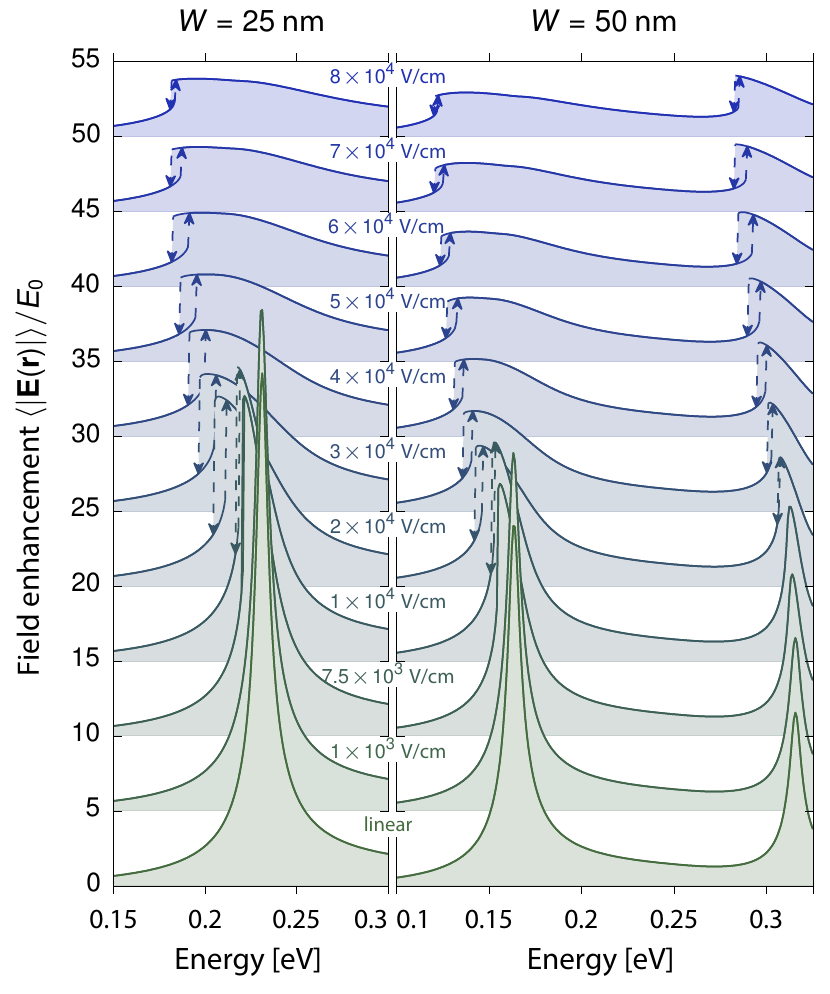}
\caption{Field enhancement $\langle|\Ev(\rv)|\rangle/E_0$ as a function of energy $\hbar\omega$, for varying incident field strengths $E_0$ (as indicated above each spectrum). Each spectrum is offset vertically by 5 units. Two ribbon widths $W=25\ \text{nm}$ and $50\ \text{nm}$ are examined. Regions of bistability are delimited by dashed arrows which indicate the ramping direction. Material parameters are as in Fig.~\ref{fig:disprel}. \label{fig:pw_spectral}}
\end{figure}
%-----------------

%--------------------------------------------------------------------------
\vskip .35em
\emph{Plane-wave excitation and bistability.}\ \ %
Having considered the dispersion of eigenmodes, we next turn our attention to the response of the system due to a normally incident plane wave, polarized along $x$, i.e. $\Ev_{\text{ext}}(z=0) = E_0\hat{\mathbf{x}}$ and $\phi_{\text{ext}}(z=0)=-E_0xW$, corresponding to vanishing $\kpar$. In addition to the power absorbed from the incident wave, the induced and total electric fields are of primary interest -- here we focus on the latter. For reasons of numerical efficiency, and as we shall see, physical necessity, we compute for each separate energy the response by an initial linear calculation, followed by a ramping of the incident field strength. Specifically, for fixed $\hbar\omega$, we consider a  ramp-array $\{E_{0,n}\}_{n=1}^{N}$ with $E_{0,n+1}>E_{0,n}$ and with $E_{0,1}$ sufficiently small to be considered a linear perturbation. Starting from $E_{0,1}$ we compute associated solutions and proceed, generally, to field strength $E_{0,n+1}$ with initial guesses on $f$ and $\phi$ obtained from the $n$th solution. This defines the upward ramp, corresponding to slowly turning the incident intensity up. Upon reaching $n=N$ we invert the procedure and follow a downward ramp, in the pattern $E_{0,n} \rightarrow E_{0,n-1}$, corresponding to slowly turning the intensity down. 

In Fig.~\ref{fig:pw_spectral} we examine the spectral response of ribbons of widths $W=25\ \text{nm}$ and $50\ \text{nm}$ under different excitation strengths, i.e.\ under varying $E_0$. For moderately high $E_0$ the linear Lorentzian resonance is asymmetrically perturbed, slightly broadened, and redshifted. Furthermore, the upward and downward ramps to $E_0$ give identical spectra. As $E_0$ is increased further, these perturbations intensify. However, in certain frequency ranges the response to upward and downward ramps toward $E_0$ differ (regions delimited by dashed arrows); a trademark of bistability. Similar features were discussed for extended graphene in Ref.~\onlinecite{Peres:2014} using the Kerr model of Eq.~\eqref{eq:Kerr} and in Ref.~\onlinecite{Cox:2014a} for finite systems using a phenomenological anharmonic model. The primary extension here is the full self-consistent accounting of the inhomogeneous nonlinear conductive profile arising in nanostructured systems. For comparison, we note that the bistability studied here is achievable at much larger energies than in the extended system, where it is restricted to $\hbar\omega<\sqrt{4/3}\alpha_{\scriptscriptstyle\text{fs}} \ef$ under normal incidence (with $\alpha_{\scriptscriptstyle\text{fs}}\equiv e^2/4\pi\varepsilon_0\hbar c$)~\cite{Peres:2014}.
Here, bistability is evident in the dipole mode for both $W=25\ \text{nm}$ and $50\ \text{nm}$, but also visible for the quadrupole mode for $W=50\ \text{nm}$. In both cases, the area traced by the bistable region initially increases with $E_0$ and then decreases due to saturation and increased absorption.  

%---------------------
\begin{figure}[!htb] 
\includegraphics{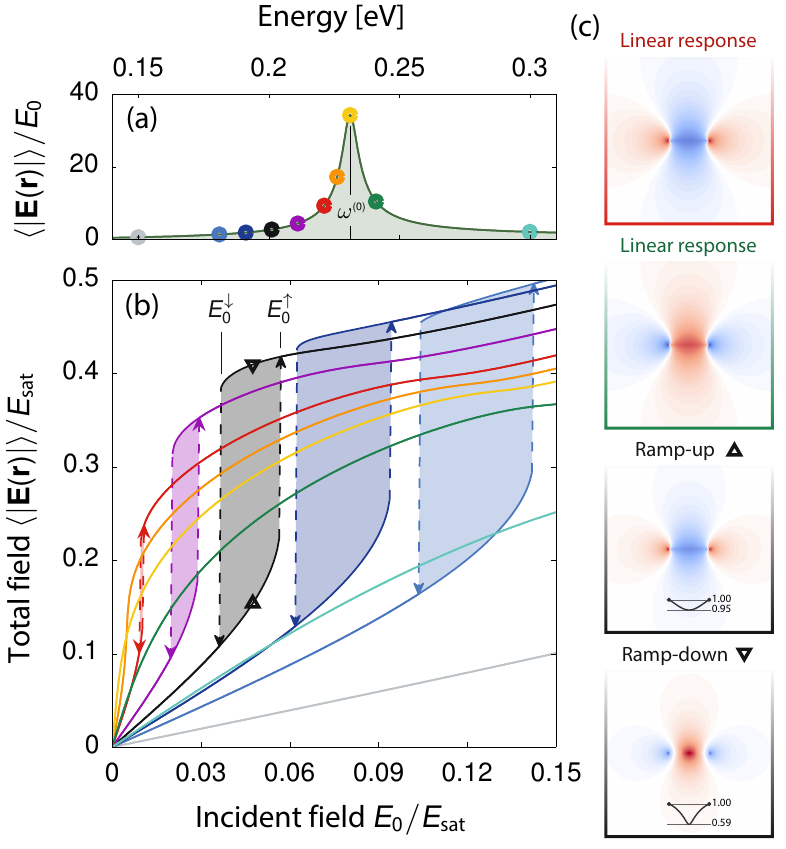}
\caption{Hysteresis arising from bistable behavior in a $W=25\ \text{nm}$ nanoribbon excited by a plane wave $E_0 \hat{\mathbf{x}}$ (material parameters are as in Fig.~\ref{fig:disprel}). \textsf{(a)}~Linear response field-enhancement spectrum versus energy. Selected energies are highlighted by colored markers, and the linear resonance energy $\hbar\omega^{\scriptscriptstyle(0)}$ is labeled explicitly. \textsf{(b)}~Hysteresis curves at fixed energy [corresponding colorwise to those highlighted in (a)] for total field $\langle|\Ev(\rv)|\rangle$ versus incident field $E_0$. Bistable regions are indicated by shading and delimited by energy-dependent low- and high-point field strengths $E_0^{\downarrow\uparrow}$. \textsf{(c)}~Intensity maps of the induced electric field $\mathrm{Re}\big[E^{\text{ind}}_x(x,z)\big]$. Colorscale is identical across the four maps, ranging from positive (red), through zero (white), to negative (blue) in a symmetric range. Absolute magnitudes are scaled logarithmically for intelligibility. Frame color indicates association with energies in (a). Field strengths in the high-field maps are specified by corresponding triangles in (b). Sparklines, defined as in Fig.~\ref{fig:disprel}(b), indicate the range and variation of $|f(\rv)|$.\label{fig:hysteresis}}
\end{figure}
%---------------------

The history dependence of the response is further examined in Fig.~\ref{fig:hysteresis}(b), depicting hysteresis curves of $E_0$ vs. $\langle|\Ev(\rv)|\rangle$ (normalized to the frequency-dependent saturation field) at a selection of fixed frequencies as indicated in the linear spectrum of Fig.~\ref{fig:hysteresis}(a). At energies far from the linear resonance at $\hbar\omega^{\scriptscriptstyle(0)}$ the response $\langle|\Ev(\rv)|\rangle$ relates linearly with $E_0$. As the energy is increased towards $\hbar\omega^{\scriptscriptstyle(0)}$, a nonlinear discrepancy develops with increasing $E_0$ which eventually gives way to a discontinuous jump at a critical field strength $E_0^{\uparrow}$, indicated for a selected energy in Fig.~\ref{fig:hysteresis}(b). As $E_0$ is reduced on the downward ramp, its response initially traces out that of the upward ramp, but departs from its upward correspondent after $E_0^{\uparrow}$ and eventually undergoes a discontinuous jump at $E_0^{\downarrow}$ after which the initial path is retraced. The hysteresis area, indicated by shaded areas, increases with positive $\omega^{\scriptscriptstyle(0)}-\omega$ (though $E_0^\uparrow$ similarly increases, delaying the onset of hysteresis), but vanishes for $\omega\gtrsim\omega^{\scriptscriptstyle(0)}$ due to the redshifting of the resonance with $E_0$. The onset of bistability is reached for incident field strengths considerably below $E_{\text{sat}}$; this fortuity is of course a result of plasmonic field-enhancement of the total field.

Lastly, we comment on the field profiles of the excitations. First we highlight the linear response at energies just below and above $\hbar\omega^{\scriptscriptstyle(0)}$, indicated by red and green markers in Fig.~\ref{fig:hysteresis}(a) and~\ref{fig:hysteresis}(c). The field profile exhibits a well-known $\pi$ phase-shift between the two energies, a result which can be appreciated e.g.\ by inspection of the linear harmonic-oscillator polarizability $\alpha(\omega) \propto [(\omega^{\scriptscriptstyle(0)})^2 - \omega(\omega+i\gamma)]^{-1}$ which exhibits a sign-change of its real part as $\omega$ traverses $\omega^{\scriptscriptstyle(0)}$: as a result, the induced dipole $p(\omega) = \alpha(\omega)E_0$ changes sign for $\omega\lessgtr\omega^{\scriptscriptstyle(0)}$, and correspondingly so for the induced fields. A similar sign change is observed in the bistable comparison, see black-framed modes in Fig.~\ref{fig:hysteresis}(c). Again, the origin of the sign change can be appreciated from a polarizability consideration by including a third-order anharmonic term to the harmonic oscillator model~\cite{Cox:2014a}, see SM.

%--------------------------------
%----- SUMMARY & DISCUSSION -----
%--------------------------------
\vskip .35em
\emph{Summary and discussion.}\ \ %
In this paper we have analyzed the impact of Kerr nonlinearity on the plasmonic response of graphene nanostructures, specifically for nanoribbons. The key distinction of nanostructures compared to the corresponding extended system arises from the strongly inhomogeneous fields of localized plasmonic excitations, which in turn incur an inhomogeneous conductive profile. We have derived a simple analytic expression, Eq.~\eqref{eq:perturb}, which approximates the nonlinear resonance shifts, while accounting for both inhomogeneity and overall amplitude of the nonlinear perturbation. The characteristic field of the Kerr nonlinearity in graphene is the saturation field $E_{\text{sat}}$. However, significant nonlinear interaction can be achieved near plasmonic resonances even for much weaker incident fields owing to plasmonic field enhancement. Finally, we discussed the existence of a plasmonic bistability in nanoribbons under normal incidence. 

The applications of optical bistabilities are well-known and long-pursued~\cite{Boyd:2008,Gibbs:1985}, with implications particularly in optical switching. Indeed, a range of platforms have been scrutinized for this purpose, in recent years e.g.\ in photonic crystal cavities (PCC) where nonlinearities are enhanced by large $Q$-factors and light-slowdown~\cite{Soljacic:2004}. Whether graphene can further the state-of-the-art in this mature field remains to be seen~\cite{Khurgin:2014}. We expect, however, that a very profitable avenue for progress exists in hybrid approaches, utilizing e.g.\ PCC and graphene in unison -- as has in fact been explored experimentally~\cite{Gu:2012}, albeit without taking advantage of the resonant plasmonic nonlinearity described herein. A simultaneous tuning of both cavity \emph{and} plasmonic resonance should allow for maximal nonlinear manifestation in such systems.
Advances in this direction requires improved understanding of nonlinearities in nanostructures; the present work constitutes one such effort. Several features, however, remain unexplored, underscoring the fertility and richness of the field. For example, from a semiclassical perspective, barring atomistic approaches~\cite{Cox:2014a,Cox:2015a}, questions remain relating to the role of interband nonlinearities~\cite{Cheng:2014}, nonlocality, and the effective role of edge states.

In closing, we mention a final question of singular practical relevance, namely damage thresholds. So far, to the best of our knowledge, measurements do not exist in the infrared, but in the optical domain~\cite{Krauss:2009,Roberts:2011,Currie:2011} the reported thresholds fall in the rather broad range from $\sim\!10^{6}\ \text{V/cm}$ in fs-pulsed operation~\cite{Roberts:2011} to just $\sim\!10^{4}\ \text{V/cm}$ for hour-long continuous wave operation~\cite{Krauss:2009}. For comparison, the saturation field at $\hbar\omega = \ef = 0.2\ \text{eV}$ is $E_{\text{sat}} \approx 6.7\times 10^{5}\ \text{V/cm}$. Though direct comparison is impossible, in part due to frequency range, pulse conditions, and the uncertain impact of field enhancement, this highlights that even resonantly enhanced nonlinearities in graphene walk a narrow road -- not unlike previous contenders for large nonlinearities. Given the promising results presented herein, however, we believe the journey will be worth the effort.

%----------------------------
%----- ACKNOWLEDGEMENTS -----
%----------------------------
\vskip 1em
\emph{Acknowledgments.}\ \ %
The Center for Nanostructured Graphene is sponsored by the Danish National Research Foundation, Project DNRF58.
This work was also supported by the Danish Council for Independent Research, Project 1323-00087.
W.Y. acknowledges support from the Lundbeck Foundation, grant no.~70802.

%----------------------
%----- REFERENCES -----
%----------------------

\bibliographystyle{apsrev4-1}
%\bibliography{../../References/Christensen_bib}

%

\newpage
% % % % % % % % % % % % % % % % % % % % % % % % % % % % % % % % % % % % % % % % % % % % % % % % % % % % % % % % % % % % % % % % % % % % % % % % % % % % % % % % % % % % % % % % %  
% % % % % % % % % % % % % % % % % % % % % % % % % % % % % % % % % % % % % % % % % % % % % % % % % % % % % % % % % % % % % % % % % % % % % % % % % % % % % % % % % % % % % % % % % 
% % % % % % % % % % % % % % % % % % % % % % % % % % % % % % % % % % % % SUPPLEMENTAL MATERIAL % % % % % % % % % % % % % % % % % % % % % % % % % % % % % % % % % % % % % % % % % % 
% % % % % % % % % % % % % % % % % % % % % % % % % % % % % % % % % % % % % % % % % % % % % % % % % % % % % % % % % % % % % % % % % % % % % % % % % % % % % % % % % % % % % % % % %
% % % % % % % % % % % % % % % % % % % % % % % % % % % % % % % % % % % % % % % % % % % % % % % % % % % % % % % % % % % % % % % % % % % % % % % % % % % % % % % % % % % % % % % % % 

%-----CHANGES TO SETUP-----

%-----APPPEND 'S' TO REFERENCES OF ALL KINDS-----
\renewcommand{\theequation}{S\arabic{equation}} %Equations
\renewcommand{\thefigure}{S\arabic{figure}} %Figures
\renewcommand{\bibnumfmt}[1]{[S#1]} %Citations
\renewcommand{\citenumfont}[1]{S#1}
\setcounter{equation}{0}
\setcounter{figure}{0}
\newgeometry{top=1in, bottom=1.5in, left=1.25in, right=1.25in}

\onecolumngrid

%-----TITLE-----
\noindent{\Large\bfseries\textsf{SUPPLEMENTAL MATERIAL\vskip 1.5em}}

%-----CHANGE SETUP FOR PARAGRAPH INDENTS AND SKIPS-----
\setlength{\parindent}{0em}
\setlength{\parskip}{.5em}

\titleformat{\section}
  {\normalfont\normalsize\bfseries\sffamily}{\thesection.}{.5em}{\MakeUppercase}
\titlespacing*{\section}{0em}{2em}{.5em}
%-----------------------
%----- MAIN MATTER -----
%-----------------------

%-------------------------------
%----- ITERATIVE PROCEDURE -----
%-------------------------------
\section{Iterative procedure for nonlinear problem}
We here discuss an iterative approach to solving the nonlinear equation
\begin{equation}\label{eq:eigsys_nonlin_driven}
\lambda\phi(\rv) = \lambda\phi_{\text{ext}}(\rv) + \textsf{V}\textsf{D}\big[f[\phi]\big]\phi(\rv),
\end{equation} 
which is essentially just the driven correspondent of Eq.~\eqref{eq:eigsys}, and where we have emphasized the dependence of $\textsf{D}$ on $\phi(\rv)$ through $f(\rv)$. The problem is evidently nonlinear, but can be solved efficiently by iteration with only linear algebra at each step. We follow the usual iteration scheme, as e.g.\ also used previously in the studies of bistability in dielectric waveguides~\cite{Wang:1995_S}:
\begin{enumerate}  \setlength{\parskip}{0pt}   \setlength{\itemsep}{2pt}  
\item Compute a linear solution based on an initial guess of $f=f_{\text{ini}}$, i.e.\ solve Eq.~\eqref{eq:eigsys_nonlin_driven} with $\textsf{D}\big[f[\phi]\big]\rightarrow\textsf{D}[f=f_{\text{ini}}]$. Denote the obtained solution as $\phi^{[0]}$. Set the iteration step $m=0$.
\item Calculate the $m$th guess at the occupation function $f^{[m]}$ from the potential $\phi^{[m]}$. \label{step:itermp1}
\item Compute the $(m+1)$th iteration by solving the linear system $\lambda\phi^{[m+1]} = \lambda\phi_{\text{ext}} + \textsf{V}\textsf{D}\big[f^{[m]}\big]\phi^{[m+1]}$.\label{step:itermp2}
\item Iterate steps~\ref{step:itermp1} and \ref{step:itermp2} until convergence, otherwise update iteration step $m\rightarrow m+1$.
\end{enumerate}
We impose convergence criteria corresponding to the simultaneous fulfillment of (with $\text{tol} = 10^{-5}$)
\begin{subequations}
\begin{align}
\max_{\rv\in\Omega}\big|\phi^{[m+1]}(\rv)-\phi^{[m]}(\rv)\big|   \big/   \max_{\rv\in\Omega}\big|\phi^{[m]}(\rv)\big|&< \text{tol},\\
\max_{\rv\in\Omega}\big|f^{[m+1]}(\rv)-f^{[m]}(\rv)\big|    \big/    \max_{\rv\in\Omega}\big|f^{[m]}(\rv)\big|&< \text{tol},
\end{align}
\end{subequations}
being of standard type for iterative approaches to nonlinearity~\cite{Wang:1995_S}. In all considered cases the iterative procedure converged after at most several hundred iterations. One exception should be mentioned however; the dipolar eigenmodes at field strengths $3\times 10^5\ \text{V/cm}$ and $3.5\times 10^5\ \text{V/cm}$ failed to converge after 1250 iterations for $\kpar \gtrsim 5$ and are consequently absent in Fig.~\ref{fig:disprel} for these momenta. This could likely be remedied by a more elaborate stepping procedure, though such investigations have not been pursued further in this work.

Two additional extensions of the simple iterative scheme described above are employed. Firstly, for numerical stability we apply a linear mixing scheme for updating guesses on $f$, specifically we use $\textsf{D}\big[f^{[m]}_{\text{mix}}\big]$ with $f^{[m]}_{\text{mix}} = (1-\xi_{\text{mix}})f^{[m-1]} + \xi_{\text{mix}}f^{[m]}$ in step~\ref{step:itermp1} (mixing parameter $\xi_{\text{mix}} = 0.275$) rather than the unmixed $\textsf{D}\big[f^{[m]}\big]$. 
Secondly, the initial guess $f_{\text{ini}}$ is always taken from the previous field strength in ramping scenarios. This provides a significant numerical speed-up and, crucially, allows us to investigate hysteresis and bistability. The initial guess at the first field strength is naturally $f_{\text{ini}}=1$. 

For eigenmodal calculations where $\phi_{\text{ext}}=0$, we normalize $\phi_n$ at each iteration to impose the desired ribbon-averaged field strength $\langle |\Ev(\rv)|\rangle$, and in addition determine $\omega_n$ from $\lambda_n(\omega_n)$ by numerically solving the equation in the complex frequency-plane.

%------------------------------------
%----- DISCRETIZATION PROCEDURE -----
%------------------------------------
\section{Matrix representation of $\mathsf{V}$ and $\mathsf{D}$ in a discretized basis}
We here elaborate the reduction of the differential and integral operators $\mathsf{D}$ and $\mathsf{V}$ to matrix representations $\textbf{\textsf{D}}$ and $\textbf{\textsf{V}}$ using an equidistant discrete basis. Specifically, we discuss the 1D ribbon case, although the generalization to general 2D restrictions is straightforward. 
Specifically, we imagine a system in the $xy$-plane, translationally invariant along $y$ and with finite extent along $x$. For simplicity, we assume just a single ribbon, such that $x$ is limited to the simple domain $x\in[0,1]$. Furthermore, as the operators necessarily act on a potential $\phi(\rv)$, we impose translational invariance along $y$ by the decomposition $\phi(\rv) = \phi(x)\e^{\mathrm{i}\kpar y}$. 

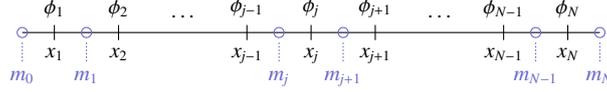
\begin{figure}
\noindent\begin{tikzpicture}
\def\ribw{3.8}
\def\tickh{.2}
\def\tickxstep{\ribw/4.5}
\def\scatext{.9}

\tikzset{dots/.style={dash pattern=on .6pt off .7pt,color=black!35!blue!60!white}}

%The "graphene-line"
\draw (-\ribw,0) -- (\ribw,0);

%"Regular" ticks and dots

\draw (-\ribw+\tickxstep/2,-\tickh/2) node [below,scale=\scatext] {$x_1$} -- (-\ribw+\tickxstep/2,\tickh/2) node [above,scale=\scatext] {$\phi_1$};
\draw (-\ribw+\tickxstep/2+\tickxstep,-\tickh/2) node [below,scale=\scatext] {$x_2$} -- (-\ribw+\tickxstep/2+\tickxstep,\tickh/2) node [above,scale=\scatext] {$\phi_2$};

\draw (-\ribw+\tickxstep/2+2*\tickxstep,\tickh/2) node [above,scale=\scatext] {$\ldots$};

\draw (-\ribw+\tickxstep/2+3*\tickxstep,-\tickh/2) node [below,scale=\scatext] {$x_{\!j-1}$} -- (-\ribw+\tickxstep/2+3*\tickxstep,\tickh/2) node [above,scale=\scatext] {$\phi_{\!j-1}$};
\draw (-\ribw+\tickxstep/2+4*\tickxstep,-\tickh/2) node [below,scale=\scatext] {$x_{\!j}$} -- (-\ribw+\tickxstep/2+4*\tickxstep,\tickh/2) node [above,scale=\scatext] {$\phi_{\!j}$};
\draw (-\ribw+\tickxstep/2+5*\tickxstep,-\tickh/2) node [below,scale=\scatext] {$x_{\!j+1}$} -- (-\ribw+\tickxstep/2+5*\tickxstep,\tickh/2) node [above,scale=\scatext] {$\phi_{\!j+1}$};

\draw (-\ribw+\tickxstep/2+6*\tickxstep,\tickh/2) node [above,scale=\scatext] {$\ldots$};

\draw (-\ribw+\tickxstep/2+7*\tickxstep,-\tickh/2) node [below,scale=\scatext] {$x_{N-1}$} -- (-\ribw+\tickxstep/2+7*\tickxstep,\tickh/2) node [above,scale=\scatext] {$\phi_{N-1}$};
\draw (-\ribw+\tickxstep/2+8*\tickxstep,-\tickh/2) node [below,scale=\scatext] {$x_N$} -- (-\ribw+\tickxstep/2+8*\tickxstep,\tickh/2) node [above,scale=\scatext] {$\phi_N$};

%Mid- and boundary points

\draw [color=black!35!blue!60!white] (-\ribw,0) circle (\tickh/3);;
\draw [dots] (-\ribw,-\tickh/1.5) -- (-\ribw,-\tickh*2) node [below,scale=\scatext,color=black!35!blue!60!white] {$m_0$};

\draw [color=black!35!blue!60!white] (-\ribw+1*\tickxstep,0) circle (\tickh/3);;
\draw [dots] (-\ribw+1*\tickxstep,-\tickh/1.5) -- (-\ribw+1*\tickxstep,-\tickh*2) node [below,scale=\scatext,color=black!35!blue!60!white] {$m_1$};

\draw [color=black!35!blue!60!white] (-\ribw+4*\tickxstep,0) circle (\tickh/3);;
\draw [dots] (-\ribw+4*\tickxstep,-\tickh/1.5) -- (-\ribw+4*\tickxstep,-\tickh*2) node [below,scale=\scatext,color=black!35!blue!60!white] {$m_{\!j}$};

\draw [color=black!35!blue!60!white] (-\ribw+5*\tickxstep,0) circle (\tickh/3);;
\draw [dots] (-\ribw+5*\tickxstep,-\tickh/1.5) -- (-\ribw+5*\tickxstep,-\tickh*2) node [below,scale=\scatext,color=black!35!blue!60!white] {$m_{\!j+1}$};

\draw [color=black!35!blue!60!white] (-\ribw+8*\tickxstep,0) circle (\tickh/3);;
\draw [dots] (-\ribw+8*\tickxstep,-\tickh/1.5) -- (-\ribw+8*\tickxstep,-\tickh*2) node [below,scale=\scatext,color=black!35!blue!60!white] {$m_{N-1}$};

\draw [color=black!35!blue!60!white] (\ribw,0) circle (\tickh/3);
\draw [dots] (\ribw,-\tickh/1.5) -- (\ribw,-\tickh*2) node [below,scale=\scatext,color=black!35!blue!60!white] {$m_N$};
\end{tikzpicture}
\caption{Sketch of the discretization approach applied to a single ribbon.\label{fig:discretization}}
\end{figure}

Starting with the differential operator $\mathsf{D}$, we consider its operation onto $\phi(\rv)$, which takes the form $\mathsf{D}\phi(\rv) = \partial_{x}[f(x)\partial_{x}\phi(x)]\e^{\mathrm{i}\kpar y} - \kpar^2f(x)\phi(x)\e^{\mathrm{i}\kpar y}$. By extension, we define the operation of $\mathsf{D}$ onto the single-variable function $\phi(x)$ through $\mathsf{D}\phi(x) \equiv \partial_{x}[f(x)\partial_{x}\phi(x)] - \kpar^2f(x)\phi(x)$.
To proceed, we introduce a discretization of the $x$-coordinates as $\{x_{\!j}\}_{\!j=1}^N$ with associated values $\phi_{\!j}\equiv\phi(x_{\!j})$ and $f_{\!j}\equiv f(x_{\!j})$ (we take $N=150$, being well-converged in all considered cases). Though not strictly necessary, we assume equidistant $x_{\!j}$ with constant spacing $x_{\!j+1}-x_{\!j} = a$, see Fig.~\ref{fig:discretization}.

The matrix elements $D_{\!jl}$ of the finite-element representation of $\mathsf{D}$ is then defined by $\mathsf{D}\phi_{\!j} = \sum_l D_{\!jl}\phi_l$. The elements can be deduced using finite differences at the midpoints. Specifically, using central differences $\partial_x[f_j\partial_x \phi_{\!j}] \simeq a^{-1}(m_{\!j}-m_{\!j-1})$ where $m_{\!j}$ defines midpoint-values of the function $m(x) \equiv f(x)\partial_x\phi(x)$ such that $m_{\!j}\simeq(2a)^{-1} (f_{\!j+1}+f_{\!j})(\phi_{\!j+1}-\phi_{\!j})$, see Fig.~\ref{fig:discretization}.
For all interior points, $j\in[2,N-1]$, this then allows a decomposition of $D_{\!jl}$ as the tridiagonal matrix
\begin{subequations}
\begin{equation}
D_{\!jl} = \tfrac{1}{2a^2}\Big[\delta_{\!j-1,l}(f_{\!j-1}+f_{\!j}) - \delta_{\!jl}(f_{\!j-1}+2f_{\!j} + f_{\!j+1})  + \delta_{\!j+1,l}(f_{\!j}+f_{\!j+1})\Big] - \delta_{\!jl}\kpar^2 f_{\!j}.
\end{equation}
At the end-points $j=1$ and $j=N$ we explicitly account for boundary conditions. Specifically, we ensure a vanishing of normal current, equivalent to the condition $\partial_x\phi(x) = 0$ for $x=0$ and $x=1$. In turn, this forces $m_0 = m_N = 0$, allowing
\begin{align}
D_{1l} &= \tfrac{1}{2a^2}(f_1+f_2)(-\delta_{1,l} + \delta_{2,l}) -\delta_{1,l}\kpar^2 f_1,\\
D_{Nl} &= \tfrac{1}{2a^2}(f_{N-1}+f_N)(\delta_{N-1,l} - \delta_{N,l}) -\delta_{N,l}\kpar^2 f_N.
\end{align}
\end{subequations}
As an alternative to taking explicit account of the boundary condition, one can allow a slightly larger $x$-range, and explicitly include points with $f(\rv)=0$ outside $\rv\in\Omega$ -- the step in $f(\rv)$ at $\rv\in\partial\Omega$ then mimics an edge charge and accounts numerically for the boundary condition; such a procedure may be preferable in finite structures without any geometric symmetries compatible with a square grid.

The integral operator $\mathsf{V}$ is similarly amenable to explicit expression on the equidistant grid. Specifically, letting $\mathsf{V}$ operate on a function $g(\rv) = g(x)\e^{\mathrm{i}\kpar y}$ one finds~\cite{GradshteynRyzhik_S,Wang:2013_S}
\begin{equation}\label{eq:CoulombMatrixKernel} 
\mathsf{V}g(\rv) = \e^{\mathrm{i}\kpar y}\!\int\!\dd{x'} 2 K_0(\kpar|x-x'|) g(x'),
\end{equation}
where $\kpar>0$ is assumed and with $K_0$ denoting the zeroth order modified Bessel function of the second kind. Assuming a slowly varying $g(x)$ and an equidistant $\{x_{\!j}\}$ then allows a matrix decomposition of $\mathsf{V}$ via $\mathsf{V}g_{\!j} = \sum_{l}V_{\!jl}g_{\!j}$ where~\cite{DLMF_S}
\begin{align} \label{eq:CoulombElements} 
V_{\!jl} &= 2 \int_{x_l-a/2}^{x_l+a/2}\!\!\dd{x'} K_0(\kpar|x_{\!j}-x'|) 
= \pi\!\!\!\!\sum_{\tilde{x}=x_{\!jl} \pm a/2}\!\!\!\!
\tilde{x}\Big\lbrace K_0(\kpar|\tilde{x}|)\Big[\mathbf{L}_1(\kpar|\tilde{x}|) + \tfrac{2}{\pi}\Big] + K_1(\kpar|\tilde{x}|)\mathbf{L}_0(\kpar|\tilde{x}|)\Big\rbrace,
\end{align}
with $x_{\!jl} \equiv x_{\!j}-x_l$ and $\mathbf{L}_{0,1}$ denoting modified Struve functions of zeroth and first order.

A final detail which should be discussed is the special case $\kpar=0$, where the kernel $K_0(\kpar|x-x'|)$ in Eq.~\eqref{eq:CoulombElements} diverges. Despite this divergence, finite and meaningful matrix elements can be retrieved by invoking charge conservation. Specifically, we note the small argument expansion $K_0(\kpar|x-x'|) \sim - \ln(|x-x'|) -\ln(\kpar) + \alpha$ where $\alpha = \ln(2)-\gamma_{\text{\textsc{em}}}$ ($\gamma_{\text{\textsc{em}}}$ is the Euler--Mascheroni constant)~\cite{DLMF_S}. The $x'$-independent term $-\ln(\kpar)+\alpha$ gives a contribution $[-\ln(\kpar)+\alpha]\int \dd{x'} g(x')$ to Eq.~\eqref{eq:CoulombMatrixKernel} and appears divergent as $\kpar\rightarrow 0$. Nevertheless, this contribution vanishes for the functions $g(\rv')$ of relevance since they always represent induced charges [as evident from Eq.~\eqref{eq:phi_integrodiff}] and obey charge conservation $\int \dd{x'} g(x')=0$. As such, the $\kpar = 0$ case can be calculated by simply letting $K_0(\kpar|x-x'|)\rightarrow -\ln(|x-x'|)$ in Eq.~\eqref{eq:CoulombElements}~\cite{Wang:2013_S}, yielding $V_{\!jl} = -2 \sum_{s=\pm} s(x_{\!jl}+s\tfrac{a}{2}) \ln(|x_{\!jl}+s\tfrac{a}{2}|)$ for $\kpar=0$.

This concludes the real-space discretization approach for reduction of the abstract operator equation of Eq.~\eqref{eq:eigsys} into a matrix equation $\lambda\bm{\phi} = \textbf{\textsf{V}}\textbf{\textsf{D}}\bm{\phi}$ with $\bm{\phi}$ denoting the vector form of $\phi_{\!j}$.

%-----------------------------------------------------------
%----- PERTURBATIVE APPROACH TO EIGENVALUE CORRECTIONS -----
%-----------------------------------------------------------
\section{Perturbation estimate of the nonlinear shift of eigenfrequencies}
We here provide the derivations that allow the approximate result of Eq.~\eqref{eq:perturb}. As we explain below, the approach relies on the formulation of a Hermitian eigenproblem followed by application of standard perturbation theory to a spatially inhomogeneous problem.

The compound operator $\textsf{V}\textsf{D}$ defined in Eq.~\eqref{eq:eigsys} is -- though numerically practical -- inconvenient for analytical considerations, because it is not symmetric. However, the problem can (of course) be cast as a Hermitian eigenproblem with eigenvalues $\lambda_n$ [though, strictly speaking, only for real, positive occupation functions $f(\rv)$, which we restrict our analysis to here], as also noted recently in Refs.~\onlinecite{Velizhanin:2015_S} and~\onlinecite{AbajoManjavacas:2015_S}. Specifically, consider the application of the scaled gradient operation $-\sqrt{f(\rv)}\nabla$ onto Eq.~\eqref{eq:phi_integrodiff}:
\begin{equation}
-\lambda\sqrt{f(\rv)}\nabla\phi(\rv) = \sqrt{f(\rv)}\nabla \int_{\Omega} \!\mathrm{d}^2{\rv'}\, V(\rv,\rv') \nabla'\!\cdot\Big\lbrace\sqrt{f(\rv')}\big[-\sqrt{f(\rv')}\nabla'\phi(\rv')\big]\Big\rbrace.
\end{equation}
Defining the scaled in-plane field $\bm{\xi}(\rv)\equiv -\sqrt{f(\rv)}\nabla\phi(\rv)$ and manipulating further allows
\begin{align}
\lambda\bm{\xi}(\rv) &= \sqrt{f(\rv)}\nabla \int_{\Omega} \!\mathrm{d}^2{\rv'}\, V(\rv,\rv') \nabla'\!\cdot\big[\sqrt{f(\rv')}\bm{\xi}(\rv')\big] 
\nonumber\\
&\overset{a}{=} \sqrt{f(\rv)}\nabla \Bigg\lbrace 
\int_{\Omega} \!\mathrm{d}^2{\rv'}\, \nabla'\!\cdot \Big[V(\rv,\rv') \sqrt{f(\rv')}\bm{\xi}(\rv')\Big] -
\int_{\Omega} \!\mathrm{d}^2{\rv'}\, \big[\nabla' V(\rv,\rv')\big]\cdot \big[ \sqrt{f(\rv')}\bm{\xi}(\rv')\big]
\Bigg\rbrace
\nonumber \\
&\overset{b}{=} -\sqrt{f(\rv)}\nabla \int_{\Omega} \!\mathrm{d}^2{\rv'}\, \sqrt{f(\rv')} \big[\nabla' V(\rv,\rv')\big]\cdot \bm{\xi}(\rv')
\nonumber \\
&\overset{c}{=} -\int_{\Omega} \!\mathrm{d}^2{\rv'}\, \sqrt{f(\rv)f(\rv')} \big[\nabla\otimes\nabla' V(\rv,\rv')\big] \bm{\xi}(\rv')
\end{align}
with associated steps $a-c$ explicated below for convenience:
\begin{enumerate}[label=$\alph*$.\ ] \setlength{\parskip}{0pt}   \setlength{\itemsep}{2pt}  
\item Application of chain rule to expand integrand.
\item The first integral term in step $a$ vanishes, as can be deduced by application of the divergence theorem which transforms the term to $\oint_{\partial\Omega} V(\rv,\rv') \sqrt{f(\rv')} \big[\bm{\xi}(\rv')\cdot \mathbf{n}'\big]$. The integrand vanishes for all $\rv'\in\partial\Omega$ due to the no-spill boundary condition on the induced current which forces $ \bm{\xi}(\rv')\cdot \mathbf{n}' = 0$ on $\rv'\in\partial\Omega$.
\item The term $\sqrt{f(\rv)}\nabla$ is taken under the integral sign. $\nabla$ operates on $\rv$ and hence only on $V(\rv,\rv')$. The operation $\nabla\big\lbrace \big[\nabla'V(\rv,\rv') \big]\cdot \bm{v}(\rv')\big\rbrace$ is rewritten in the equivalent outer-product form $\big[\nabla\otimes\nabla' V(\rv,\rv')\big]\bm{v}(\rv')$ with elements $[\nabla\otimes\nabla']_{ij} = \partial_{r_i}\partial_{r_{\!j}'}$. 
\end{enumerate}
We then define the operator $\textsf{M}$ by its action on a field-ket $|\bm{\xi}\rangle$ [where, as usual, $\langle \rv|\bm{\xi}\rangle \equiv \bm{\xi}(\rv)$]
\begin{equation}\label{eq:defMoper}
\langle \rv|\textsf{M}|\bm{\xi}\rangle \equiv \int_{\Omega} \!\mathrm{d}^2{\rv'}\, \sqrt{f(\rv)f(\rv')} \big[\nabla\otimes\nabla' V(\rv,\rv')\big]\bm{\xi}(\rv'),
\end{equation}
with associated eigenspectrum $(-\lambda_n)$ and $|\bm{\xi}_n\rangle$:
\begin{equation}
(-\lambda_n)|\bm{\xi}_n\rangle  = \textsf{M} |\bm{\xi}_n\rangle.
\end{equation}
The operator $\textsf{M}$ is evidently symmetric, positive semi-definite, and thus Hermitian. Aaccordingly, the eigenspectrum $\{-\lambda_n\}$ is non-negative and real; and the eigenkets $|\bm{\xi}_n\rangle$ are orthogonal $\langle \bm{\xi}_n|\bm{\xi}_{n'}\rangle = \delta_{nn'} \langle \bm{\xi}_n|\bm{\xi}_{n}\rangle$ and span the solution space for $\rv\in\Omega$.

With these facts established, we can now discuss a perturbation treatment. Specifically, we consider the simple case where $f(\rv) = f^{\scriptscriptstyle (0)} + \delta f^{\scriptscriptstyle (1)}(\rv)$ for $\rv\in\Omega$ with ``groundstate'' $f^0 = 1$ and perturbation $f^1$ with strength $\delta$. The corresponding expansion of $\textsf{M} = \textsf{M}^{\scriptscriptstyle (0)} + \delta \textsf{M}^{\scriptscriptstyle (1)} + \mathcal{O}(\delta^2)$ is found by expansion of Eq.~\eqref{eq:defMoper}, yielding
\begin{subequations}\label{eq:perturbM}
\begin{align}
\langle\mathbf{\rv}|\textsf{M}^{\scriptscriptstyle (0)}|\bm{\xi}\rangle &=  \int_{\Omega} \!\mathrm{d}^2{\rv'}\, 
\big[\nabla\otimes\nabla' V(\rv,\rv')\big]\bm{\xi}(\rv'), \label{eq:perturbM0}\\ 
\langle\mathbf{\rv}|\textsf{M}^{\scriptscriptstyle (1)}|\bm{\xi}\rangle &= \frac{1}{2}\int_{\Omega} \!\mathrm{d}^2{\rv'}\,
\big[ f^{\scriptscriptstyle (1)}(\rv) + f^{\scriptscriptstyle (1)}(\rv') \big]
\big[\nabla\otimes\nabla' V(\rv,\rv')\big]\bm{\xi}(\rv').\label{eq:perturbM1}
\end{align}
\end{subequations}
Since $\textsf{M}$ is a Hermitian operator usual perturbation theory applies~\cite{Gasiorowicz:2003_S}. Specifically, for a ``groundstate'' eigenspectrum $\{-\lambda_n^{\scriptscriptstyle (0)},|\bm{\xi}^{\scriptscriptstyle (0)}_n\rangle\}$ the leading-order correction to the perturbed eigenvalue $\lambda_n = \lambda_n^{\scriptscriptstyle (0)} + \delta \lambda_n^{\scriptscriptstyle (1)} + \mathcal{O}(\delta^2)$ is derivable by application of Eqs.~\eqref{eq:perturbM} [by using the $(\rv,\rv')$-symmetry of the resulting equation]
\begin{equation}
\lambda^{\scriptscriptstyle (1)}_n = -\frac{ \langle \bm{\xi}_n^{\scriptscriptstyle (0)} | \textsf{M}_1 | \bm{\xi}_n^{\scriptscriptstyle (0)} \rangle }
{\langle \bm{\xi}_n^{\scriptscriptstyle (0)} | \bm{\xi}_n^{\scriptscriptstyle (0)} \rangle}
\nonumber\\
= \lambda_n^{\scriptscriptstyle (0)} \frac{ \langle \bm{\xi}_n^{\scriptscriptstyle (0)} | f^{\scriptscriptstyle (1)} | \bm{\xi}_n^{\scriptscriptstyle (0)} \rangle }
{\langle \bm{\xi}_n^{\scriptscriptstyle (0)} | \bm{\xi}_n^{\scriptscriptstyle (0)} \rangle}.
\end{equation}

For nonlinear purposes, we unfortunately do not know the exact perturbation $f^{\scriptscriptstyle (1)}$ as it should be determined self-consistently with the total field $|\bm{\xi}_n\rangle$. However, for low field-strengths this self-consistency can be neglected and we can approximate $f[|\bm{\xi}_n\rangle] \simeq f[|\bm{\xi}_n^{\scriptscriptstyle (0)}\rangle]$ with $|\bm{\xi}_n^{\scriptscriptstyle (0)}\rangle$ referring to the electric field predicted by a \emph{linear} calculation (at the desired field strength). For the Kerr-type nonlinearity of Eq.~\eqref{eq:Kerr} the resulting correction is therefore [assuming vanishingly small loss and noting $\bm{\xi}^{\scriptscriptstyle (0)}(\rv) = \Ev^{\scriptscriptstyle (0)}(\rv)$ for $f^{\scriptscriptstyle (0)}=1$]
\begin{equation}
\lambda_n^{\scriptscriptstyle (1)} \simeq  -\lambda_n^{\scriptscriptstyle (0)} \frac{9}{8}
\frac{ \int_{\Omega} \!\mathrm{d}^2{\rv}\, |\mathbf{E}^{\scriptscriptstyle(0)}(\rv)|^4}{E_{\text{sat}}^2 \int_{\Omega} \!\mathrm{d}^2{\rv}\, |\mathbf{E}^{\scriptscriptstyle(0)}(\rv)|^2} = 
-\lambda_n^{\scriptscriptstyle (0)} \frac{9}{8}
\frac{ \langle |\mathbf{E}^{\scriptscriptstyle(0)}(\rv)|^4\rangle }{E_{\text{sat}}^2 \langle |\mathbf{E}^{\scriptscriptstyle(0)}(\rv)|^2\rangle},
\end{equation}
with $E_{\text{sat}}$ similarly evaluated at the linear resonance frequency $\omega_n^{\scriptscriptstyle (0)}$ associated with $\lambda_n^{\scriptscriptstyle (0)}$. Finally, the result of the main text, Eq.~\eqref{eq:perturb}, is obtained by invoking the relation between eigenvalues $\lambda_n$ and eigenfrequencies $\omega_n$ together with the lossless intraband conductivity $\sigma_{\scriptscriptstyle(1)}(\omega) \simeq \mathrm{i}e^2\ef/\pi\hbar^2\omega$.

%---------------------------------------
%----- ANHARMONIC OSCILLATOR MODEL -----
%---------------------------------------
\section{Qualitative anharmonic oscillator model}
We review the basics of the simple anharmonic oscillator model~\cite{Boyd:2008_S,Cox:2014a_S}, and discuss how it -- in connection with a polarizability consideration -- explains the $\pi$ phase-shift observed for the bistable solutions in Fig.~\ref{fig:hysteresis}(c). 

In this qualitative model, we represent the induced dipole by a single (time-dependent) coordinate $\textsf{x}$, which obeys the simple equation of motion
\begin{equation}\label{eq:effectivenewton}
m \ddot{\textsf{x}} + m\gamma\dot{\textsf{x}} = -efE_0(t) - \partial_{\textsf{x}}U(\textsf{x}),
\end{equation}
with an effective anharmonic restoring potential $U(\textsf{x}) = \tfrac{1}{2}m(\omega^{\scriptscriptstyle(0)})^2 \textsf{x}^2 - \tfrac{1}{4}m a \textsf{x}^4$, effective oscillator mass $m$, linear resonance $\omega^{\scriptscriptstyle(0)}$, anharmonic parameter $a$ (note that $a>0$ in our case cf.\ sign of Kerr conductivity), and coupling factor $f$. We seek the solution that oscillates at $\e^{-\mathrm{i}\omega t}$ in response to a perturbation $E_0(t)=E_0(\omega)\e^{-\mathrm{i}\omega t}$, i.e.\ the Kerr response; we denote this term by $\textsf{x}^{\scriptscriptstyle(1\omega)}(\omega)\e^{-\mathrm{i}\omega t}$. Working with Eq.~\eqref{eq:effectivenewton} one finds (omitting declaration of $\omega$-dependence)
\begin{equation}
m\Big[(\omega^{\scriptscriptstyle(0)})^2-\omega(\omega+\mathrm{i}\gamma) - 3 a|\textsf{x}^{\scriptscriptstyle(1\omega)}|^2 \Big] \textsf{x}^{\scriptscriptstyle(1\omega)} = -e f E_0. 
\end{equation}
The polarizability $\alpha^{\scriptscriptstyle(1)}$ is linked to $\textsf{x}^{\scriptscriptstyle (1\omega)}$ via the induced dipole $p^{\scriptscriptstyle (1\omega)} = -e \textsf{x}^{\scriptscriptstyle (1\omega)} = \alpha^{\scriptscriptstyle (1\omega)} E_0$, allowing (ignoring loss, being nonessential for the present considerations)
\begin{equation}\label{eq:alphaanharmon}
\Big[ (\omega^{\scriptscriptstyle(0)})^2 - \omega^2 - 3a e^{-2} |\alpha^{\scriptscriptstyle(1\omega)}|^2 E_0^2 \Big] 
\alpha^{\scriptscriptstyle(1\omega)} = e^2f/m.
\end{equation}
For the bistable scenarios the term $(\omega^{\scriptscriptstyle(0)})^2-\omega^2$ is always positive, see e.g.\ Figs.~\ref{fig:pw_spectral} and~\ref{fig:hysteresis}. Depending on the magnitude of $3ae^{-2}\alpha^{\scriptscriptstyle(1\omega)}E_0^2$ relative to $(\omega^{\scriptscriptstyle(0)})^2-\omega^2$ it is then clear that polarizability-solutions of opposing sign can arise, depending on the sign of the terms bracketed on the left-hand side of Eq.~\eqref{eq:alphaanharmon}. Furthermore, if we denote the positive and negative solutions $\alpha^{\scriptscriptstyle(1\omega)}_{+}$ and $\alpha^{\scriptscriptstyle(1\omega)}_{-}$, respectively, it can then be deduced by direct inspection of Eq.~\eqref{eq:alphaanharmon} that $|\alpha^{\scriptscriptstyle(1\omega)}_{+}|<|\alpha^{\scriptscriptstyle(1\omega)}_{-}|$. In other words, the induced dipole -- and hence the induced fields -- of the positive solution should be lower than its negative counterpart; upon identifying the lower branches of Fig.~\ref{fig:hysteresis}(b) with $\alpha_{+}^{\scriptscriptstyle(1)}$ and vice versa for the upper branch, we see that this is exactly the case. As such, the anharmonic model describes not only the phase-shift, but also the magnitude interrelationship. Lastly, we mention for completeness that the anharmonic model describes also a third solution, which, however, is physically irrelevant as it is unstable (and correspondingly is not found in the iterative procedure employed in this study, nor in experimental investigation).

%--------------------
%-----REFERENCES-----
%--------------------

\bibliographystyle{apsrev4-1}

\end{document}